# TESTS OF THE HUBBLE LAW

# FROM THE LUMINOSITY FUNCTION

# OF IRAS GALAXIES

Michael A. Strauss

*Institute for Advanced Study, Princeton, New Jersey*

Daniel M. Koranyi

*Mathematics Department, Princeton University*

**Abstract:** All direct measurements of peculiar velocities of galaxies assume the Hubble law at low redshifts. However, it has been suggested by Segal *et al.* (1993) that the correlation of redshifts and fluxes in a complete sample of *IRAS* galaxies is inconsistent with the Hubble law, and instead implies that redshifts are proportional to the square of distance at small redshift. The essence of the argument is that the luminosity function of galaxies itself can be used as a distance indicator. We examine this possibility critically using a variety of statistical tests, and conclude that the data do indeed support the Hubble law, although the strongest arguments require the assumption of small density contrasts on large scales. We identify several possible pitfalls in such an analysis which could lead one to an erroneous conclusion.

---

Peculiar velocities of galaxies are defined observationally as the difference between the observed redshift of a galaxy, and some measure of its distance. This definition explicitly assumes the Hubble Law, which states that in the absence of peculiar velocities, redshifts and distances of galaxies are strictly proportional at distances small relative to the Hubble radius. The observational evidence for this is based on identifying a "standard candle", a subset of the galaxies whose luminosity is found to be constant, or to depend in a calibratable way on measurable parameters (Peebles 1993). Luminous radio galaxies seem to be a standard candle in their K-band light (Lilly & Longair 1982), the optical and infrared luminosity of spiral galaxies can be predicted from their rotational width (Tully & Fisher 1977), and the luminosity of brightest cluster galaxies can be predicted from the slope of their surface brightness profiles (Lauer & Postman 1992). These and other studies indicate strongly that distances are indeed proportional to redshifts, as the Hubble law states.

Rather than use individual galaxies as standard candles, one might take the approach of assuming that averaged over large enough scales, the galaxy population as a whole is independent of position, in which case one might use the galaxy luminosity function as a standard candle. In this case, given the luminosity function, one can make predictions for the distribution of fluxes in a galaxy sample as a function of redshift, predictions that will presumably be different given different models for the relation between redshift and distance. Segal *et al.* (1993) have analyzed the *IRAS* galaxy redshift survey of Strauss *et al.* (1992) with this in mind, and claim that the data are not consistent with the Hubble Law, but rather with redshifts of galaxies being proportional to the *square* of their distances from us.

We test these claims with the redshift survey of *IRAS* galaxies by Fisher (1992), consisting of all *IRAS* galaxies over 88% of the sky with 60 micron fluxes above 1.2 Jy (5319 galaxies in all).

We calculate luminosity functions from the sample for a given distance-redshift relation using the maximum-likelihood method discussed in Yahil *et al.* (1991) and references therein. This method gives a luminosity function independent of density inhomogeneities within the volume spanned by the survey. We take two approaches: *a)* Use the parameterized form of the luminosity function given by Yahil *et al.*, maximizing the likelihood with respect to these parameters; *b)* Use the non-parametric (step-wise) method of Nicoll & Segal (1982) and Efstathiou *et al.* (1988), maximizing the likelihood with respect to the value of the luminosity function at a series of discrete steps. Figures 1 and 2 show the resulting luminosity function for distance-redshift relations given by $z \propto d^p$, for $p = 1, 2$, and 3. The two methods give results in very good agreement with one another, especially for $p = 2$ and $p = 3$. The discontinuities inherent in the luminosity function in the non-parametric approach introduce systematic effects in the statistics we calculate below, thus we restrict ourselves to the parametric approach for the moment.

If the galaxy population as a whole is uniform on large scales, the luminosity function determined from subsamples of the data set at different redshifts should agree. This is tested in Figure 3 for $p = 1$ and $p = 2$. The dashed lines in the figures are the luminosity functions determined from the full *IRAS* sample, while the solid segments are the luminosity functions determined from subsets of the data from various redshift ranges. For $p = 1$, the results for the full sample, and for its subsets, are in essentially perfect agreement, while there are severe disagreements for $p = 2$.

As mentioned above, the luminosity function is derived in a way independent of density inhomogeneities. If one does assume that when averaged over the 11.06 st of the *IRAS* survey, that the galaxy distribution is homogeneous, one can predict the distribution of redshifts with distance, given the luminosity function and a value for $p$. These predictions for various $p$ are shown in Figure 4, together with the observed distribution. The $p = 1$ curve agrees closely with that observed, although it is not perfect. One sees the effects of the Local Supercluster at $cz \approx 1500$ km s$^{-1}$, and the Great Attractor and Pisces-Perseus regions at $cz \approx 4500$ km s$^{-1}$. On very large scales, the $p = 1$ curve slightly overestimates the observed distribution, probably because of incompleteness of the sample at very high redshift (Fisher *et al.* 1992). However, if either the $p = 2$ or $p = 3$ model is correct, one would have to argue that as one looks more distant into the universe, the inhomogeneities grow. The galaxy density (ratio of observed and predicted curves) would need to be a strong function of distance from us: low nearby, rising rapidly to a maximum at 5000 — 10,000 km s$^{-1}$, and then dropping precipitously thereafter. This goes against the cosmological principle, and indeed, if such massive structures were common in the universe, we would see them reflected in the angular correlation function of faint galaxies (e.g., Maddox et al. 1990).

However, the arguments presented by Figures 3 and 4 depend on the assumption of homogeneity in the large: Figure 3 because the luminosity function of each of the different subsamples is normalized within its own volume. Is it possible to distinguish between values of $p$ without any assumptions about the galaxy distribution? Segal et al. (1993) argue that the distribution of observed fluxes in a sample in a given redshift interval is indeed independent of the galaxy distribution, and can be used to distinguish models. Figure 5 shows a closely related statistic: the distribution of luminosities in redshift intervals, for $p = 1$ and $p = 2$. In each panel, the histogram shows the observed distribution of galaxy luminosities, and the smooth curve shows the distribution predicted given the luminosity function (cf., Yahil et al. 1991). The agreement between the two appears remarkable, for both cosmologies. One can quantify the difference between the curves using a $\chi^2$ statistic, using Poisson error bars. The results are shown in each panel; in fact, $p = 1$ is shown to be superior to $p = 2$, with the latter showing $\chi^2$ systematically higher than the number of degrees of freedom in the lower redshift bins. It is surprising that the difference between $p = 1$ and $p = 2$ is so subtle. This is a consequence of two effects: $a$) In a flux-limited sample, the majority of galaxies are very close to the flux limit, thus the mean luminosity in any relatively narrow redshift interval is determined $b$) The luminosity function of the sample for different values of $p$ has very similar slope, as Figures 1 and 2 show.

The main argument in favor of the $p = 2$ model given by Segal et al. (1993) is that it correctly predicts the mean apparent magnitude in redshift bins. This comparison is shown in Figure 6, which shows the distribution of log fluxes as a function of redshift for the *IRAS* sample, together with the mean value in logarithmically spaced bins (points with errors). The two curves are the predictions of $p = 1$ and $p = 2$, now using the stepwise luminosity functions of Figure 2. First, note how close the two predictions are to one another; this is not a statistic with much power to distinguish the models! Neither statistic fits the data well at large redshifts, and both give an acceptably large value of $\chi^2$ for the 18 degrees of freedom (although $p = 1$ is worse). However, close inspection shows that the model predictions are not monotonic! How can this be? The problem has to do with the stepwise nature of the luminosity function. Figure 7 shows the resulting selection function, which enters directly into the model predictions (Figure 6). The scalloping is an artifact due to the stepwise nature of the luminosity function. We are in the process of generalizing the stepwise luminosity function method to correct this artifact, and suspect strongly that when this is fixed, the $p = 1$ model will agree better with the observed distribution of mean log flux as a function of redshift, as in Figure 6. This problem plagues Segal et al.'s analysis as well, and may be the reason they find that the $p = 2$ model is favored.

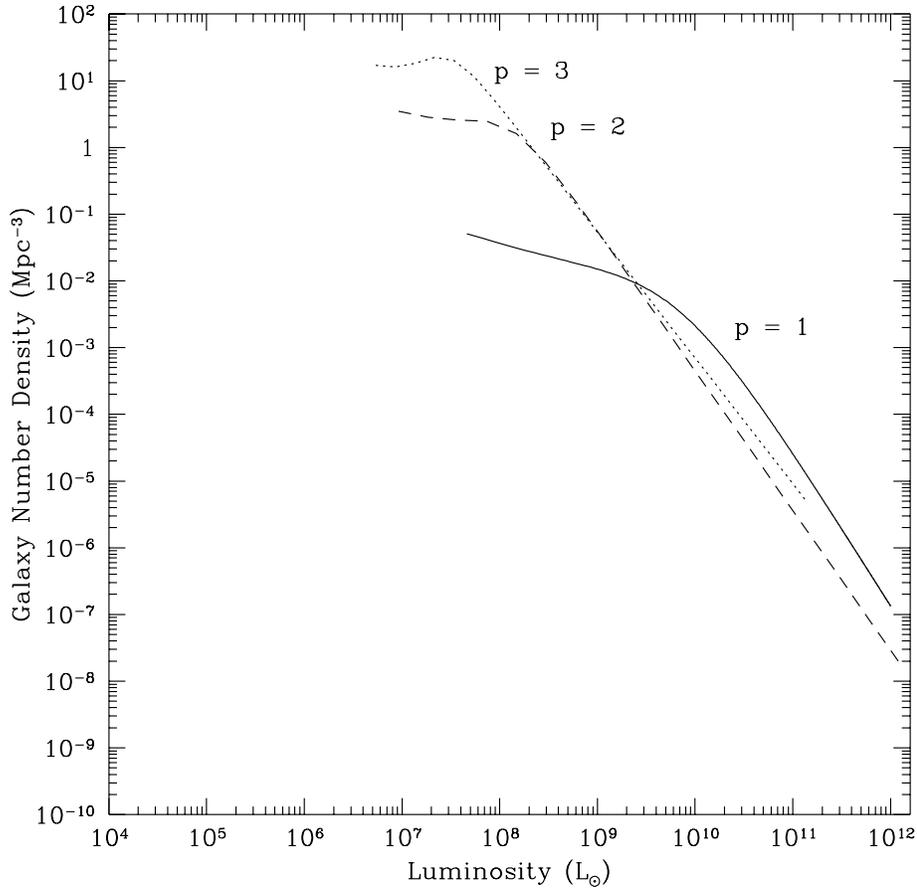

**Fig. 1** The luminosity function $\hat{\Phi}$ using a maximum likelihood fit of a parameterized form to the *IRAS* 1.2 Jy data, under the assumption of linear, quadratic, and cubic velocity-distance laws.

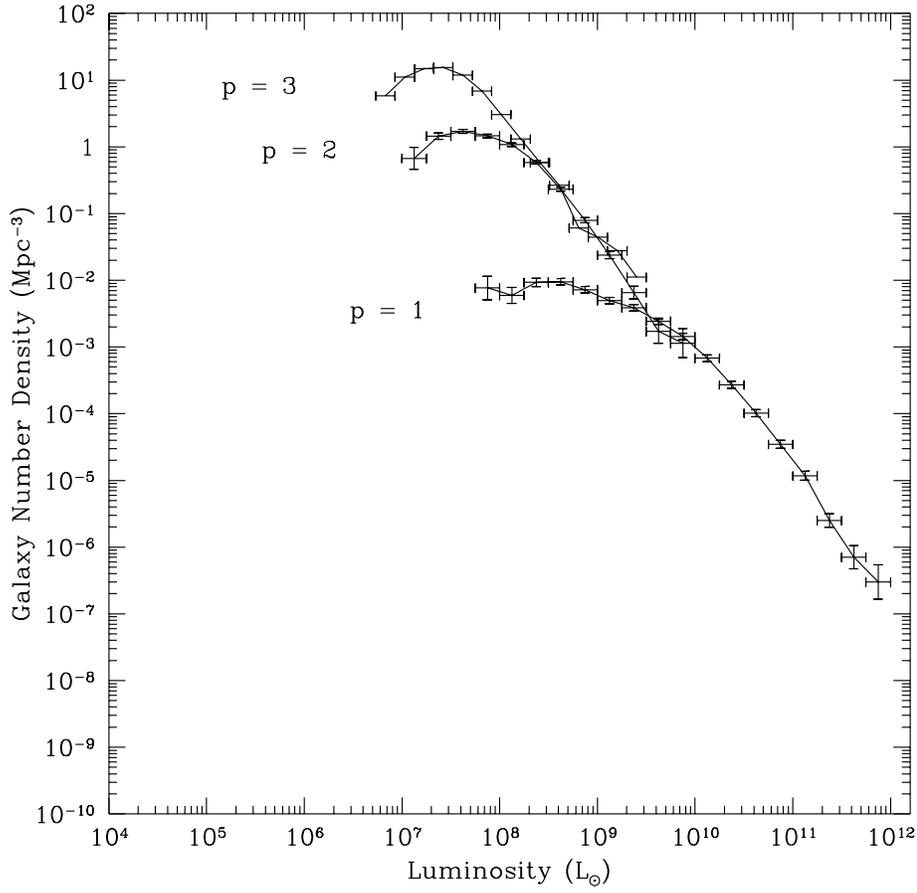

**Fig. 2** The luminosity function $\hat{\Phi}$ as determined by the iterative stepwise maximum-likelihood method, for the cases $p = 1$, $p = 2$ and $p = 3$. The errors on the $p = 3$ curve are not visible at this scale. The bin centers have been connected for ease of tracing the LF's; it is important to keep in mind that the LF's are the steps themselves.

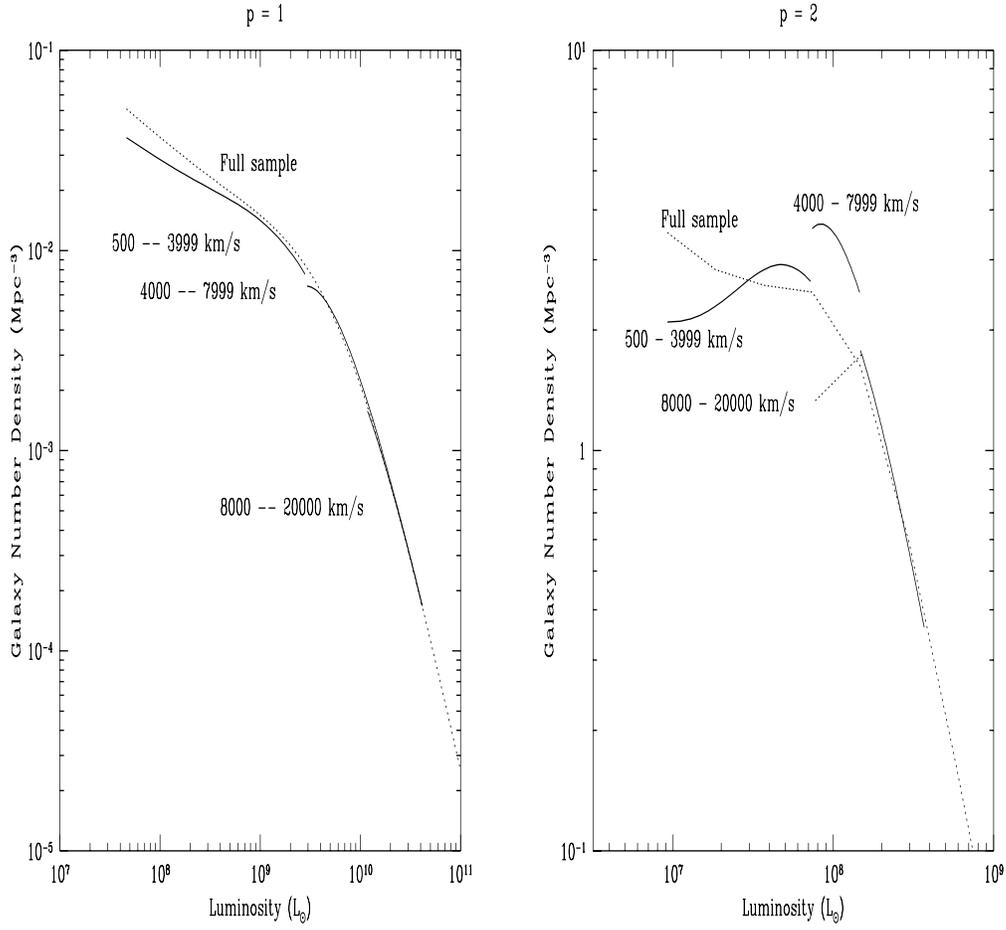

**Fig. 3** The parametrized luminosity function as given by disjoint subsamples in velocity for $p = 1$ and $p = 2$. The LF as determined by the full sample is shown in a lighter line weight for comparison.

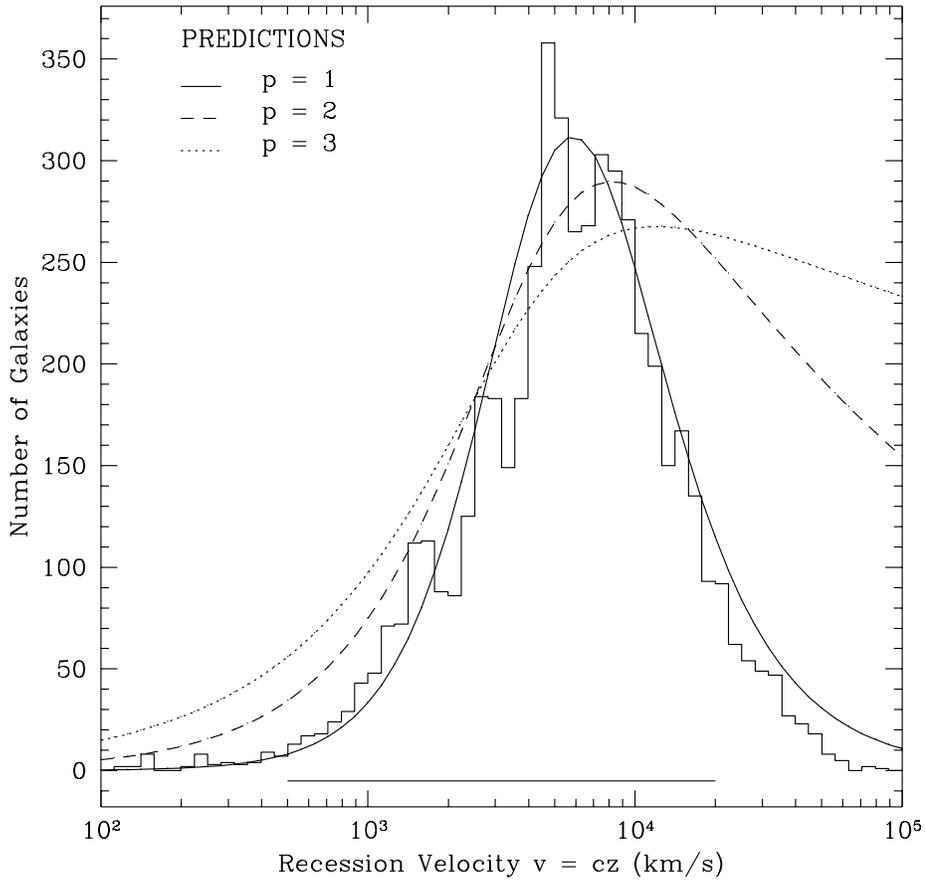

**Fig. 4** Expected *vs.* observed galaxy number counts. Histogram is for the observed data, smooth curves are predictions of cosmologies $p = 1, 2, 3$, as derived from a Yahil-parametrized LF. Isolated overdensities are due to known inhomogeneities caused by various clusters and superclusters. The bar at the bottom of the graph indicates the range (500 – 20000 km/s) over which the fit to the LF was performed.

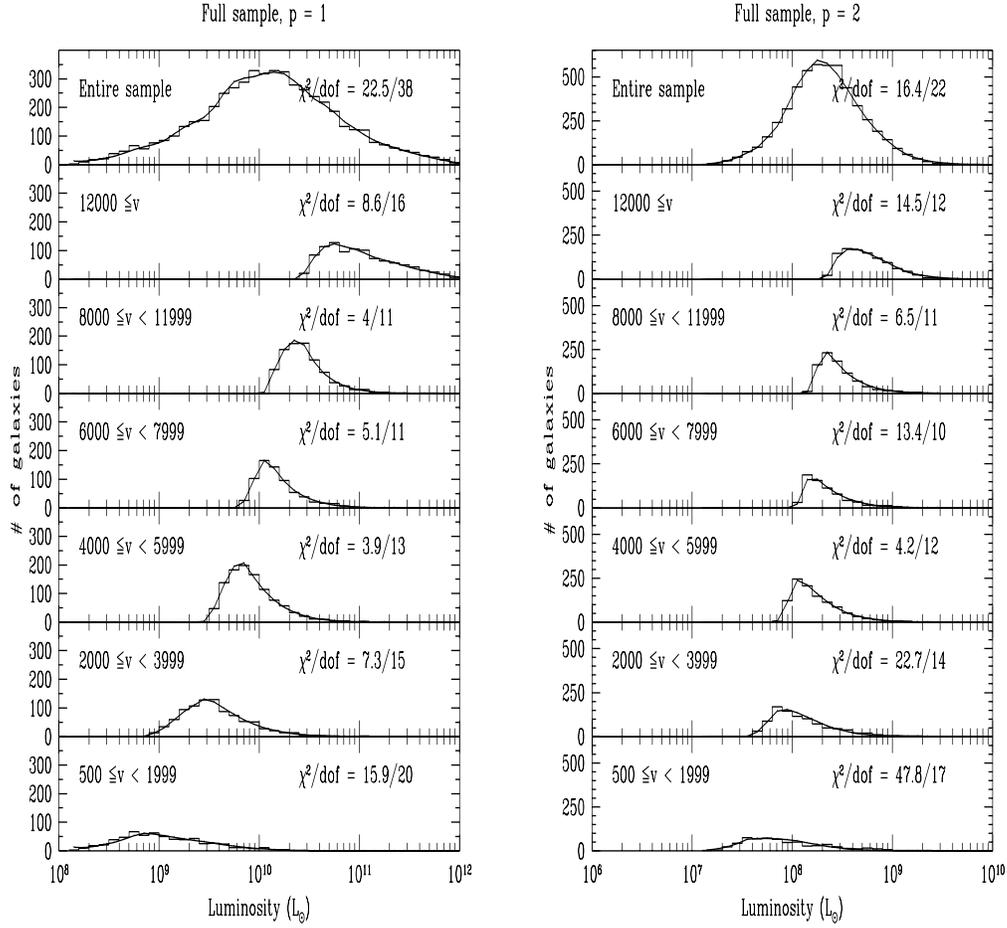

Fig. 5 Observed vs. expected distribution of galaxies in luminosity at different distance ranges, for $p = 1$ and 2. Topmost box in each case is for the entire sample in velocity. Predictions are based on a parameterized LF determined over 500 - 20000 km/s.

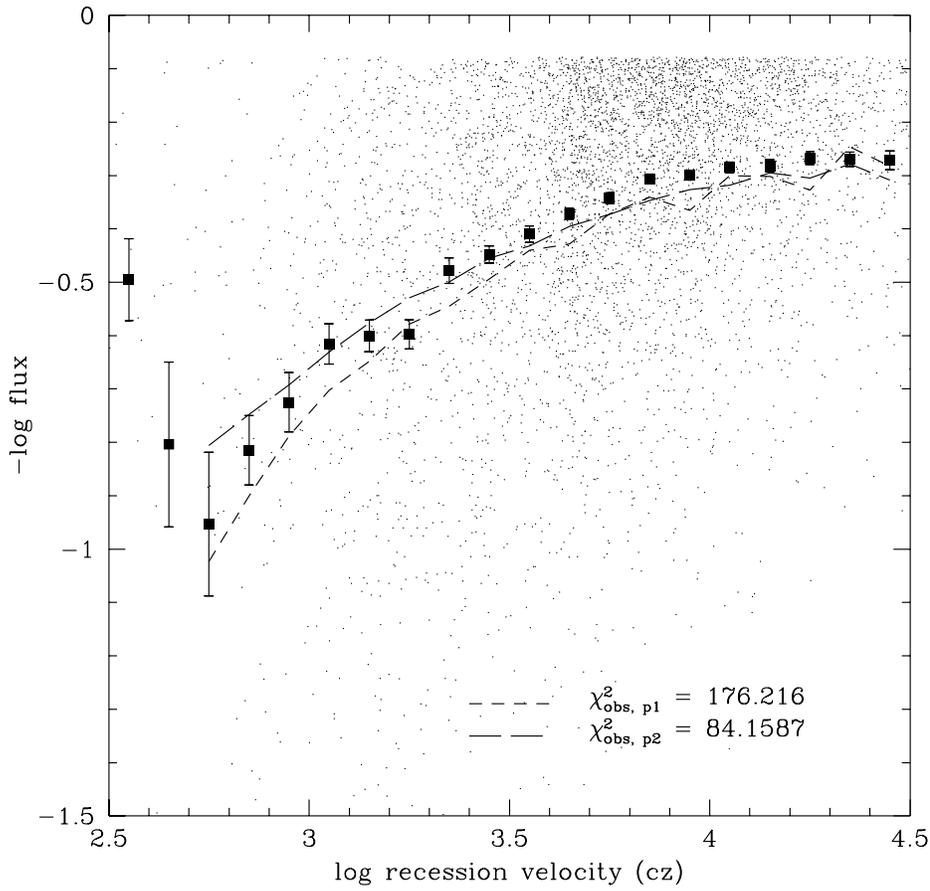

**Fig. 6** Plot of $-\log f$ as a function of $\log cz$. The points are the means in evenly spaced bins of $\log cz$; the two dashed lines are predictions for the $p = 1$ and $p = 2$ cosmologies in each redshift bin. Note that the predictions are not monotonic.

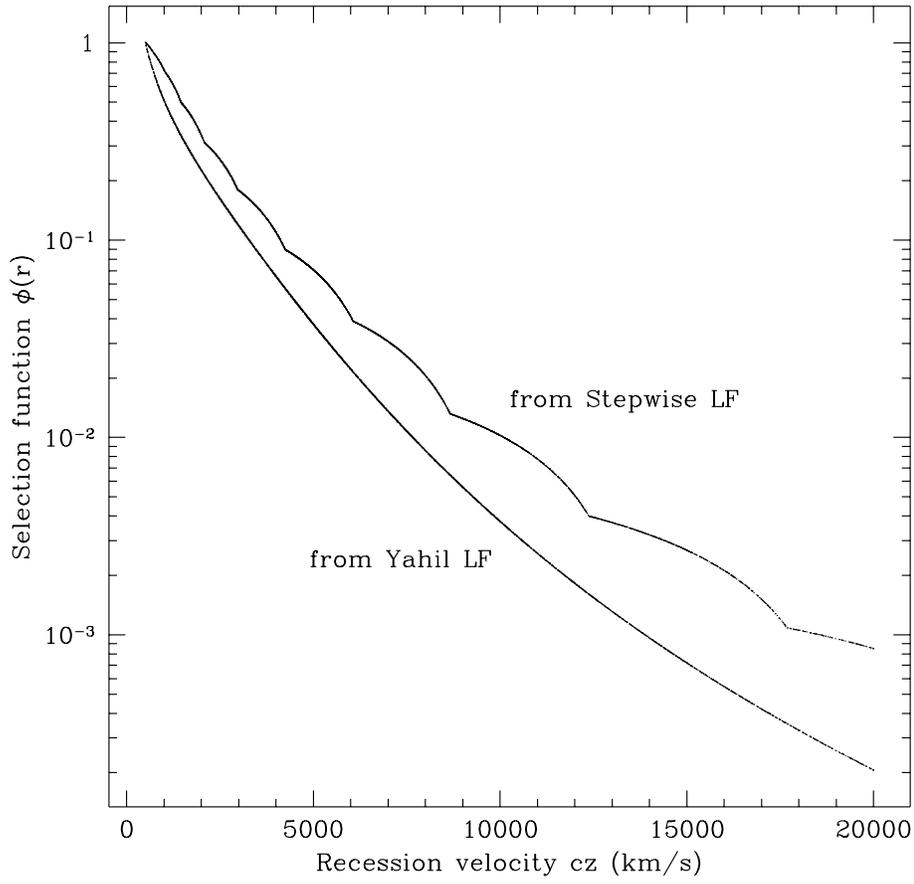

**Fig. 7** Comparison of the selection function $\phi(r)$ as rendered by the parameterized form and the stepwise LF for the case $p = 1$. The scalloped shape of the stepwise curve is an artifact of the steps; were one to interpolate the LF linearly between the step values this effect would vanish.